\documentstyle[12pt,epsfig,axodraw,a4]{article} 
\def\bild#1#2{    
        \vspace*{-5mm}
        \begin{center}
        \begin{math}
        \epsfxsize#2cm
        \epsffile{#1}
        \end{math}
        \end{center}  }
\newcommand{\vs}{\vspace{-0.25cm}}
\begin{document} 
\begin{center}
\large{\bf Chiral 2\begin{boldmath}$\pi$\end{boldmath}-exchange NN-potentials:
Two-loop contributions}

\bigskip

N. Kaiser\\

\medskip

{\small Physik Department T39, Technische Universit\"{a}t M\"{u}nchen,
    D-85747 Garching, Germany}

\end{center}

\medskip

\begin{abstract}
We calculate in heavy baryon chiral perturbation theory the local NN-potentials
generated by the two-pion exchange diagrams at two-loop order. We give explicit
expressions for the mass-spectra (or imaginary parts) of the corresponding 
isoscalar and  isovector central, spin-spin and tensor NN-amplitudes. We find 
from two-loop two-pion exchange a sizeable isoscalar central repulsion which 
amounts to $62.3\,$MeV at $r=1.0\,$fm. There is a similarly strong isovector
central attraction which however originates mainly from the third order low 
energy constants $\bar d_j$ entering the chiral $\pi N$-scattering amplitude.
We also evaluate the one-loop $2\pi$-exchange diagram with two second order 
chiral $\pi \pi NN$-vertices proportional to the low energy constants $c_{1,2,
3,4}$ as well
as the first relativistic $1/M$-correction to the $2\pi$-exchange diagrams with
one such vertex. The diagrammatic results presented here are relevant
components of the chiral NN-potential at next-to-next-to-next-to-leading order.

\end{abstract}


\bigskip
%To be published in {\it The Physical Review C (2001)}
\bigskip

Over the past years effective field theory methods have been successfully
applied to the two-nucleon system at low and intermediate energies 
\cite{weinb,kolck,kaplan,epelb,nnpap1,nnpap2}. The idea of constructing
the NN-potential from effective field theory was put forward by Weinberg
\cite{weinb} and this was first taken up by van Kolck and collaborators 
\cite{kolck} who used "old-fashioned" time-ordered perturbation theory. Later,
the systematic method of unitary transformations was employed by Epelbaum, 
Gl\"ockle and Mei{\ss}ner \cite{epelb} to construct an energy-independent 
NN-potential from the effective chiral pion-nucleon Lagrangian at 
next-to-leading order. Based on one- and two-pion exchange and nine adjustable 
NN-contact interactions which contribute only to S- and P-waves a good 
description of the deuteron properties as well as the NN phase-shifts and
mixing angles below $T_{lab}=300$\,MeV was found in that framework 
\cite{epelb}. Also recently, the elastic proton-proton scattering data base 
below $350\,$MeV laboratory kinetic energy (consisting of 1951 data points) 
has been analyzed in terms of $1\pi$-exchange and chiral $2\pi$-exchange at 
next-to-next-to-leading order in ref.\cite{nijmeg}. The resulting good
$\chi^2/dof\leq 1$ constitutes a convincing proof for the presence of the
chiral $2\pi$-exchange in the long-range proton-proton strong interaction. It
was concluded in ref.\cite{nijmeg} that $1\pi$-exchange together with chiral
$2\pi$-exchange gives a very good NN-force at least as far inwards as
$r=1.4\,$fm internucleon distance. All shorter range components of the 
NN-interaction have been effectively parametrized in ref.\cite{nijmeg} by 23 
boundary condition parameters.

At present, there is much interest in extending the calculations of the 
two-nucleon  system \cite{mach} (and also the analysis of the NN-data base) to
one higher order in the chiral expansion. This requires the full knowledge of
the next-to-next-to-next-to-leading order (N$^3$LO) chiral NN-potential. In 
momentum space this corresponds to terms in the NN T-matrix which are of fourth
order in small external momenta and the pion mass, denoted generically by 
${\cal O}(Q^4)$. In particular, all two-loop diagrams of the process $NN\to NN$
with leading order vertices contribute at that order, ${\cal O}(Q^4)$. 
In three recent publications \cite{3pi} we have calculated completely the 
static NN-potentials generated by the (two-loop) $3\pi$-exchange diagrams with 
all possible interaction vertices taken from the leading-order effective chiral
$\pi N$-Lagrangian. In these works we have evaluated the mass-spectra or
imaginary parts from which one can also easily reconstruct the momentum space
NN-amplitudes in the form of a subtracted dispersion relation. The appearing 
subtraction constants can be absorbed in the strengths of some local NN-contact
interactions which will be treated as adjustable parameters in a fit to 
NN-phase shifts etc. The purpose of this work is to present analogous results 
for the two-loop contributions to the chiral $2\pi$-exchange which also belong
to the chiral NN-potential at order ${\cal O}(Q^4)$. Furthermore, we evaluate
several new contributions  at the same order ${\cal O}(Q^4)$. These are the 
one-loop $2\pi$-exchange diagram with two second order chiral 
$\pi\pi NN$-vertices proportional to the low energy constants $c_{1,2,3,4}$ and
the first relativistic $1/M$-corrections to the $2\pi$-exchange diagrams with 
one such vertex proportional to $c_{1,2,3,4}$.  

Let us first give some basic definitions in order to fix our notation. In the
static limit and considering only irreducible diagrams the on-shell NN T-matrix
takes the following form
\begin{eqnarray} {\cal T}_{NN} &=& V_C(q) + V_S(q)\,\vec \sigma_1\cdot \vec 
\sigma_2 + V_T(q) \,\vec \sigma_1\cdot \vec q \,\,  \vec \sigma_2 \cdot \vec q
\nonumber \\ && + \Big[ W_C(q) + W_S(q)\,\vec \sigma_1\cdot  \vec \sigma_2
+ W_T(q) \,\vec \sigma_1\cdot \vec q \,\,  \vec \sigma_2 \cdot \vec q \, \Big] 
\,\vec \tau_1 \cdot \vec \tau_2\,, \end{eqnarray} 
where $q=|\vec q\,|$ denotes the momentum transfer between the initial and
final-state nucleon. The subscripts $C,S$ and $T$ refer to the central, 
spin-spin and tensor components, each of which occurs in an isoscalar
($V_{C,S,T}$) and an isovector version ($W_{C,S,T}$). As indicated, the (real) 
NN-amplitudes $V_C(q),\dots, W_T(q)$ depend only on the momentum transfer $q$ 
in the static limit $M\to \infty$, where $M$ denotes the nucleon mass. We are
interested here only in non-polynomial or finite range terms. For this purpose 
it is sufficient to calculate the imaginary parts of the NN-amplitudes, 
Im\,$V_{C,S,T} (i\mu)$ and Im\,$W_{C,S,T}(i\mu)$, which result from analytical 
continuation to time-like momentum transfer $q=i\mu-0^+$ with $\mu\geq 2m_\pi$.
These imaginary parts are also the mass-spectra entering a representation of
the local coordinate-space potentials in the form of a continuous superposition
of Yukawa-functions,     
\begin{eqnarray} \widetilde V_C(r) &=& -{1\over 2\pi^2 r} \int_{2m_\pi}^\infty
d\mu \,\mu \,e^{-\mu r} \, {\rm Im}\, V_C(i\mu)\,, \\ \widetilde V_S(r) &=& 
{1\over 6\pi^2 r} \int_{2m_\pi}^\infty d\mu \,\mu \,e^{-\mu r} \Big[ \mu^2\,
{\rm Im}\, V_T(i\mu) - 3\, {\rm Im}\,V_S(i\mu) \Big]\,, \\ \widetilde V_T(r)
&=& {1\over 6\pi^2 r^3} \int_{2m_\pi}^\infty d\mu\,\mu\, e^{-\mu r}(3+3\mu r+
\mu^2r^2){\rm Im}\,V_T(i\mu)\,.\end{eqnarray}
The isoscalar central, spin-spin and tensor potentials, denoted here by
$\widetilde V_{C,S,T}(r)$, are as usual those ones which are accompanied by the
operators $1$, $\vec \sigma_1\cdot \vec \sigma_2$ and  $3\,\vec \sigma_1\cdot
\hat r\,\vec \sigma_2 \cdot \hat r -\vec \sigma_1\cdot \vec \sigma_2$,
respectively. For the isovector potentials $\widetilde W_{C,S,T}(r)$ a
completely analogous representation holds, of course.

\bigskip

\bigskip

\bild{2pi2loop.epsi}{16}
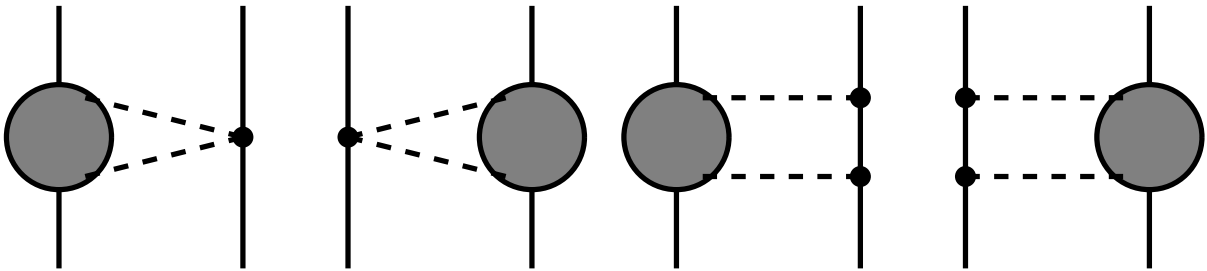
{\it Fig.1: $2\pi$-exchange diagrams at two-loop order. The grey disc
symbolizes all one-loop diagrams of elastic $\pi N$-scattering. The
combinatoric factor of the first two diagrams is 1/2. Diagrams for which
the role of both nucleons is interchanged lead to the same NN-potential.}

\bigskip

\bigskip

The two-loop $2\pi$-exchange diagrams are symbolically represented in Fig.\,1.
The grey disc should be interpreted such that it includes all one-loop diagrams
of elastic $\pi N$-scattering. The imaginary parts entering eqs.(2,3,4) can be 
calculated from these $2\pi$-exchange diagrams as integrals of the $\bar
NN\to2\pi \to \bar NN$ transition amplitudes over the Lorentz-invariant
$2\pi$-phase space making use of the Cutkosky cutting rule. In the present case
these transition amplitudes are products of tree-level $\pi N$-amplitudes and 
one-loop $\pi N$-amplitudes. The pertinent $2\pi$-phase space integral is most
conveniently performed in the $\pi\pi$ center-of-mass frame where it becomes 
proportional to a simple angular integral $\int_{-1}^1 dx$ . In the 
center-of-mass frame the on-shell pion four-momenta read $k_{1,2}^\nu = (\mu/2,
\pm \vec k\,)$ and as explained in ref.\cite{3pi} the four-momenta of the 
external nucleons can be chosen as $P_{1,2}^\nu = (\pm \mu/2,p\,\vec v\,)$ with
$\vec v$ a real unit vector and $p=\sqrt{\mu^2/4-M^2} = i\,M +{\cal O}(M^{-1})$
in the heavy nucleon limit. As an extensive check on these techniques we have 
recalculated the imaginary parts of the one-loop $2\pi$-exchange diagrams 
(considering graphs with no, single and double $\Delta$-excitations, etc.) and
we reproduced indeed exactly the results of refs.\cite{nnpap1,nnpap2} in a 
rather short calculation. 

Let us now turn to the results for the $2\pi$-exchange at two-loop order. From
the last two diagrams in Fig.\,1 we obtain the following imaginary part for 
the  isoscalar central NN-amplitude,   
\begin{equation} {\rm Im}\, V_C(i\mu) = {3g_A^4 (\mu^2-2m_\pi^2) \over \pi \mu
(4f_\pi)^6} \bigg\{ (m_\pi^2-2\mu^2) \bigg[ 2m_\pi +{2m_\pi^2 -\mu^2 \over
2\mu} \ln{\mu+2m_\pi \over \mu-2m_\pi} \bigg] +4g_A^2 m_\pi(2m_\pi^2-\mu^2)
\bigg\} \,. \end{equation}
Note that the expression in the curly bracket of eq.(5) is proportional to the
isospin-even non-spin-flip one-loop $\pi N$-amplitude Re\,$g^+_{loop}(0,\mu^2)$
(see appendix A in ref.\cite{aspects}). Modulo the last term  proportional to 
$g_A^2$ this is precisely the result of the so-called correlated $2\pi
$-exchange given in section 4 of ref.\cite{nnpap2}. Of very similar structure
is the imaginary part of the isovector spin-spin (and tensor) NN-amplitude, 
\begin{equation} {\rm Im}\, W_S(i\mu) =\mu^2 \,{\rm Im}\, W_T(i\mu) = 
{g_A^4(\mu^2-4m_\pi^2) \over \pi (4f_\pi)^6} \bigg\{ \bigg( m_\pi^2 -{\mu^2 
\over 4} \bigg)\ln{\mu+2m_\pi \over \mu-2m_\pi} +(1+2g_A^2)\mu  m_\pi\bigg\}
\,.  \end{equation}
Again, the expression in the curly bracket of eq.(6) is proportional to the
isospin-odd spin-flip one-loop $\pi N$-amplitude Re\,$h^-_{loop}(0,\mu^2)$ 
\cite{aspects}. Besides the last $g_A^2$-term,  eq.(6) agrees with result of 
correlated $2\pi$-exchange given in ref.\cite{nnpap2}. Next, we come to the 
imaginary parts of the isoscalar spin-spin and tensor NN-amplitudes. We obtain 
for them the following representation, 
\begin{eqnarray} {\rm Im}\, V_S(i\mu) =\mu^2 \,{\rm Im}\, V_T(i\mu) &=& 
{2g_A^6\mu k^3 \over (8\pi f_\pi^2)^3} \int_0^1 dx(1-x^2)\bigg\{ {48\pi^2
f_\pi^2 \over g_A^4}(d_{14}^r(\lambda)-d_{15}^r(\lambda))- \ln{m_\pi \over 
\lambda}   \nonumber \\ && -{1\over 6}+{m_\pi^2 \over k^2x^2} -\Big( 1+{m_\pi^2
\over k^2x^2} \Big)^{3/2} \ln{ k x +\sqrt{m_\pi^2 +k^2 x^2}\over  m_\pi}\bigg\}
\,,  \end{eqnarray}
with $k = \sqrt{\mu^2/4-m_\pi^2}$ the pion center-of-mass momentum. Modulo the
low-energy constant $d_{14}^r(\lambda)- d_{15}^r(\lambda)$ introduced in
ref.\cite{nadja} the expression in the curly bracket of eq.(7) is proportional
to the isospin-even spin-flip one-loop $\pi N$-amplitude Re\,$[h^+_{loop}(
\omega, \mu^2)/\omega]$  evaluated at $\omega = i\, kx$. All four diagrams in
Fig.\,1 contribute to the isovector central NN-amplitude $W_C(q)$ and we find
for its imaginary part the following representation,  
\begin{eqnarray} {\rm Im}\, W_C(i\mu)&=& {2k \over 3\mu (8\pi f_\pi^2)^3} 
\int_0^1 dx\, \Big[ g_A^2(2m_\pi^2-\mu^2) +2(g_A^2-1)k^2x^2 \Big] \bigg\{ 
3k^2x^2 \bigg( 2\ln{m_\pi \over \lambda}-1 \bigg)\nonumber \\ && +6 k x
\sqrt{m_\pi^2 +k^2 x^2} \ln{ k x +\sqrt{m_\pi^2 +k^2 x^2}\over  m_\pi}
+g_A^4(\mu^2 -2k^2 x^2 -2m_\pi^2) \nonumber \\ && \times \bigg[ {5\over 6}
-\ln{m_\pi \over \lambda} +{m_\pi^2\over k^2 x^2} -\Big( 1 +{m_\pi^2\over k^2
x^2} \Big)^{3/2} \ln{ k x +\sqrt{m_\pi^2 +k^2 x^2}\over  m_\pi} \bigg] 
\nonumber \\&& + \Big[ 4m_\pi^2 (1+2g_A^2) -\mu^2(1+5g_A^2)\Big] 
{k\over \mu} \ln {\mu +2k\over 2m_\pi} \nonumber \\ && -{\mu^2 \over 2}
(1+5g_A^2 ) \ln{m_\pi \over \lambda}+{\mu^2 \over 12} (5+13g_A^2) -2m_\pi^2 
(1+2g_A^2) \nonumber \\&& +96 \pi^2 f_\pi^2 \Big[ (2m_\pi^2-\mu^2)(d_1^r(
\lambda) +d_2^r(\lambda)) -2k^2x^2 d_3^r( \lambda)+4m_\pi^2 d_5^r(\lambda)
\Big] \bigg\}\,.   \end{eqnarray}
Again, modulo the low energy constants $d^r_j(\lambda)$ the expression in the 
curly bracket of eq.(8) is proportional to the isospin-odd non-spin-flip 
one-loop $\pi N$-amplitude Re\,$[g^-_{loop}(\omega,\mu^2)/\omega]$  evaluated 
at $\omega = i\, kx$. Furthermore, when restricting the second factor in eq.(8)
to the terms in fourth and fifth line one recovers the result of correlated 
$2\pi$-exchange given in section 4 of ref.\cite{nnpap2}. The last line in 
eq.(8) gives the contribution of the third order low energy constants 
$d^r_{1,2,3,5}(\lambda)$ introduced in ref.\cite{nadja}. Such counterterms are 
necessary in order to absorb the divergences generated by the one-loop graphs 
of elastic $\pi N$-scattering. In fact, the chiral logarithms $\ln(m_\pi 
/\lambda) $ and the scale dependent low energy constants $d_j^r(\lambda)$ in 
eqs.(7,8) combine to the (scale independent) barred quantities $\bar d_j$  
defined in ref.\cite{nadja}, or equivalently $\bar d_j=d_j^r(m_\pi)$. This 
completes the presentation of analytical results for the two-loop 
$2\pi$-exchange NN-interaction. Modulo polynomials the momentum space 
NN-amplitudes $V_C(q), \dots, W_T(q)$ can be obtained in the form of a 
subtracted dispersion relation 
\begin{equation} V_{C,S}(q) = -{2 q^6 \over \pi} \int_{2m_\pi}^\infty d\mu \,
{{\rm Im\,}V_{C,S}(i \mu) \over \mu^5 (\mu^2+q^2) }\,, \qquad  V_T(q) = {2 q^4
\over \pi} \int_{2m_\pi}^\infty d\mu \,{{\rm Im\,}V_T(i \mu) \over \mu^3
(\mu^2+q^2) }\,, \end{equation}
For the isovector amplitudes $W_{C,S,T}(q)$ a completely analogous dispersion
relation representation holds, of course. 

%\medskip

\begin{table}[hbt]
\begin{center}
\begin{tabular}{|c|ccccccccc|}
    \hline

    $r$~[fm]&0.8&0.9&1.0&1.1&1.2&1.3 &1.4& 1.5 & 1.6\\ \hline
$\widetilde V_C$~[MeV]
& 249.6 &120.3&62.28& 34.18 & 19.67& 11.77& 7.29 & 4.64 & 3.03\\ 
 $\widetilde W_S$~[MeV]
& -27.80 &-13.39&-6.93 &-3.79& -2.18&-1.30&-0.803 & -0.510 & -0.331\\ 
 $\widetilde W_T$~[MeV]
& 26.29 &12.52 &6.40& 3.47&1.97& 1.16& 0.709 & 0.445 & 0.286\\     
 $\widetilde V_S$~[MeV]
& 166.9 &71.07&32.96& 16.36&8.59&4.73& 2.71 & 1.60 & 0.977 \\ 
$\widetilde V_T$~[MeV]
& -140.0 &-59.68&-27.49&-13.55& -7.07&-3.86&-2.19 & -1.29 & -0.780\\ 
 $\widetilde W_C$~[MeV]
& -503.6 &-211.4&-96.81&-47.54&-24.73&-13.49& -7.67 & -4.51 & -2.74 \\  
  \hline
  \end{tabular}
\end{center}
%\medskip 
{\it Tab.1: Numerical values of the local NN-potentials generated by 
two-loop chiral $2\pi$-exchange versus the nucleon distance $r$. The units of 
these potentials are MeV.}
\end{table}

%\medskip

In Table\,1, we present numerical values for the coordinate space potentials
generated by chiral $2\pi$-exchange at two-loop order inserting eqs.(5,6,7,8)
into eqs.(2,3,4). We use the parameters $f_\pi = 92.4\,$MeV, $m_\pi=138\,$MeV
(average pion mass) and $g_A=1.3$. Via the Goldberger-Treiman relation this
corresponds to a strong $\pi NN$-coupling constant of $g_{\pi N}=g_A M/f_\pi=
13.2$ which is in agreement with present empirical determinations of $g_{\pi
N}$ from $\pi N$-dispersion relation analyses \cite{pavan}.  For the third
order low energy constants $\bar d_j$ we use the average values of three fits
to $\pi N$-phase shift solutions given in ref.\cite{nadja}: $\bar d_1+\bar
d_2=3.0\,$GeV$^{-2}$, $\bar d_3= -3.0\, $GeV$^{-2}$, $\bar d_5=0.1\,$GeV$^{-2}
$, $\bar d_{14}-\bar d_{15}=-5.7\, $GeV$^{-2}$. One observes from Table\,1 that
the isoscalar central potential $\widetilde V_C(r)$ generated by two-loop
$2\pi$-exchange is sizeable and repulsive. This repulsive potential is almost a
factor 100 larger than the one from the so-called correlated $2\pi$-exchange
investigated in section 4 of ref.\cite{nnpap2}. The origin of this strong
enhancement is the additional and obviously dominant $g_A^2$-term in the curly
bracket of eq.(5) which stems from all those two-loop diagrams which were not
considered in ref.\cite{nnpap2}. Similar features hold for the attractive
isovector spin-spin and the repulsive isovector tensor potentials  $\widetilde
W_{S,T}(r)$ which are however typically a factor 10 smaller than isoscalar
central potential $\widetilde V_C(r)$. At first sight, the attractive isovector
central potential $\widetilde W_C(r)$ from two-loop $2\pi$-exchange seems to be
even larger in magnitude. However, in this case the effect comes mainly from
the low energy constants $\bar d_j$. For example, if one omits the $d_j^r(
\lambda)$ in eq.(8) and sets the scale $\lambda= m_\omega =782\,$MeV in the
chiral logarithm $\ln(m_\pi/\lambda)$ the isoscalar central potential
$\widetilde W_C(r)$ gets reduced by  more than one order of magnitude. The same
phenomenon is observed for the  isoscalar spin-spin and tensor potentials
$\widetilde V_{S,T}(r)$ which is also dominated by the contribution from the
low-energy constant $\bar d_{14}-\bar d_{15}$.  In the light of these findings
one can expect that the effect of including the two-loop $2\pi$-exchange in
NN-phase shift calculations is not just a small correction. Of course, a
firm conclusion about the size of such corrections can only be drawn from the
complete N$^3$LO chiral NN-potential.

\bigskip

\bild{cifig.epsi}{14}
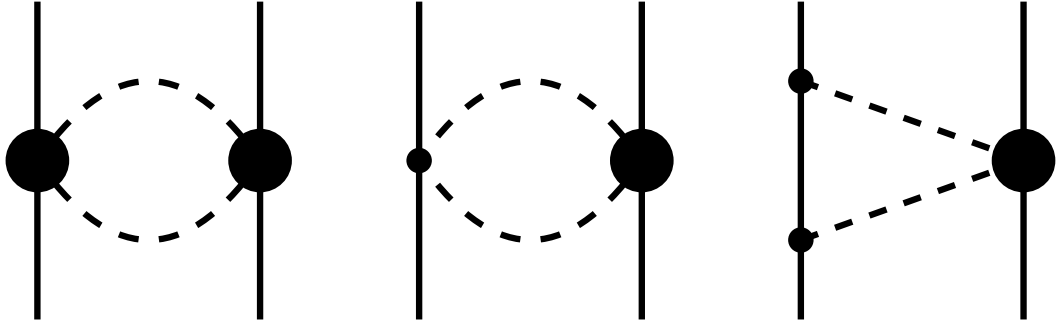
{\it Fig.2: The heavy dot symbolizes the second order chiral $\pi\pi N
N$-contact vertex proportional to the low energy constants $c_{1,2,3,4}$. The 
$2\pi$-exchange diagrams for which the role of both nucleons is interchanged 
are not shown.}

\bigskip

Furthermore, we like to present here results for another class of new 
contributions to the NN-potential at order ${\cal O}(Q^4)$. This includes the 
one-loop  $2\pi$-exchange diagram with two second order chiral $\pi\pi NN
$-vertices proportional to the low energy constants $c_{1,2,3,4}$ and the first
relativistic $1/M$-corrections to the $2\pi$-exchange diagrams with one such
vertex proportional to $c_{1,2,3,4}$.  The corresponding graphs are shown in
Fig.\,2. The first "bubble" or  "football" diagram can be straightforwardly 
evaluated in the heavy baryon formalism and one finds \cite{evgeni}, 
\begin{equation}  V_C(q) = -{3 L(q) \over 16 \pi^2 f_\pi^4 } \bigg\{
\Big[ {c_2 \over 6} w^2 +c_3 (2m_\pi^2 +q^2) -4c_1 m_\pi^2 \Big]^2 +{c_2^2
\over 45 } w^4 \bigg\}\,, \end{equation}
\begin{equation} W_T(q) = -{1\over q^2} W_S(q) = -{c_4^2\over 96 \pi^2 
f_\pi^4 } \, w^2 L(q) \,, \end{equation}
with the frequently occurring loop function,
\begin{equation} L(q) = {w\over q} \ln{w+q \over 2m_\pi} \,, \qquad w =
\sqrt{4m_\pi^2 +q^2} \,. \end{equation}
In eqs.(10,11) and in the following we omit purely polynomial terms of the 
form $const \,q^4+ const \,q^2 +const$ in the central NN-amplitudes and $const 
\,q^2 + const$ in the tensor and spin-orbit NN-amplitudes. The third diagram 
in Fig.\,2 with one $c_{1,2,3,4}$-vertex starts to contribute at order ${\cal 
O}(Q^3)$ (see section 4 in ref.\cite{nnpap1}) and the next higher order 
contribution at ${\cal O}(Q^4)$ arises from the first relativistic $1/M
$-corrections to the vertices and nucleon propagator. In order to be specific
we employ here the minimal relativistic second order $\pi N$-Lagrangian ${\cal 
L}_{\pi N}^{(2,min)}$ defined in ref.\cite{nadja}. A convenient way to obtain 
the $1/M$-corrections is to use relativistic nucleon propagators and
interaction vertices and to perform (in the NN center-of-mass frame) the 
$1/M$-expansion inside the  pion-loop integral. Consider now the second diagram
in Fig.\,2 (and its mirror partner). Since the left hand side Tomozawa-Weinberg
vertex is of isovector nature only the isovectorial $c_4$-vertex can contribute
on the right hand side. One finds modulo polynomials,   
\begin{equation} W_{SO}(q) = 2W_T(q) = -{2\over q^2} W_S(q) = -{2\over q^2}
W_C(q) = -{c_4\over 96 \pi^2 M  f_\pi^4 } \, w^2 L(q) \,. \end{equation}
As a genuine relativistic effect we encounter here a spin-orbit NN-amplitude
$W_{SO}(q)$. According to the convention of ref.\cite{nnpap1,nnpap2} the
spin-orbit amplitudes are accompanied in the NN T-matrix by the
(momentum space) spin-orbit operator  $i(\vec \sigma_1+\vec \sigma_2) \cdot 
(\vec q \times \vec p\,)$, where $\vec p$ denotes the nucleon center-of-mass 
momentum. The $c_4$-vertex in the third diagram of Fig.\,2 (and its mirror
partner) generates further isovector NN-amplitudes,  
\begin{equation} W_T(q) = -{1\over q^2} W_S(q) = {c_4 g_A^2\over 192 \pi^2 M 
f_\pi^4 } (16m_\pi^2+7q^2) L(q) \,, \end{equation}
\begin{equation} W_{SO}(q) = -{2\over q^2} W_C(q) = -{c_4 g_A^2\over 96 \pi^2 
M f_\pi^4 } (8m_\pi^2+5q^2) L(q) \,. \end{equation}
Finally, the isoscalar NN-amplitudes stem from the isoscalar $c_{1,2,3}$-vertex
in the last diagram of Fig\,.2 and they read (modulo polynomials), 
\begin{equation} V_{SO}(q) =  {c_2 g_A^2\over 16 \pi^2 M f_\pi^4 } \, w^2 L(q) 
\,, \end{equation}
\begin{equation} V_C(q) =  {g_A^2\, L(q) \over 32 \pi^2 M f_\pi^4 } \Big[ 
(c_2-6c_3) q^4 +4(6c_1+c_2-3c_3)q^2 m_\pi^2 +6(c_2-2c_3)m_\pi^4
+24(2c_1+c_3)m_\pi^6 w^{-2} \Big] \,.\end{equation}
Note that there are no isoscalar spin-spin and tensor NN-amplitudes $V_{S,T}
(q)$. This follows from the fact that to first order in the $1/M$-expansion the
$c_{1,2,3}$-vertex remains spin-independent.  For the sake of completeness we
give also the spectral representation of the spin-orbit potential in coordinate
space (accompanying the standard spin-orbit operator $-{i\over 2} (\vec
\sigma_1+\vec \sigma_2) \cdot (\vec r \times \vec \nabla)$) which reads, 
\begin{equation} \widetilde V_{SO}(r) = {1\over \pi^2 r^3} \int_{2m_\pi}^\infty
d\mu \,\mu \,e^{-\mu r} (1+\mu r)\, {\rm Im}\, V_{SO}(i\mu)\,. \end{equation}
With the help of the formula Im\,$L(i\mu)= -\pi k/\mu = -(\pi/2\mu)\sqrt{\mu^2
-4m_\pi^2}$ the numerical evaluation of the coordinate space NN-potentials
following from the one-loop results eqs.(10--17) is straightforward.     

In summary, we have calculated in this work the chiral $2\pi$-exchange
NN-potentials at two-loop order. We find that these two-loop diagrams lead to 
a sizeable isoscalar central repulsion. Furthermore, we have evaluated here the
one-loop $2\pi$-exchange diagram with two second order $c_{1,2,3,4}$-vertices 
as well as the first relativistic $1/M$-correction to the diagrams with one
such vertex. The analytical results presented here are in a form such that they
can be easily implemented in a N$^3$LO calculation of the two-nucleon system or
in an empirical analysis of low-energy elastic NN-scattering.


\vspace{-0.2cm}

\begin{thebibliography}{99} 
\bibitem{weinb} S. Weinberg, {\it Nucl. Phys.} {\bf B363}, 3 (1991).\vs
\bibitem{kolck} C. Ordonez, L. Ray and U. van Kolck, {\it Phys. Rev.} {\bf
C53}, 2086 (1996).\vs
\bibitem{kaplan} D.B. Kaplan, M.J. Savage and M.B. Wise, {\it Nucl. Phys.} {\bf
B534}, 329 (1998).\vs
\bibitem{epelb} E. Epelbaum, W. Gl\"ockle and Ulf-G. Mei{\ss}ner, {\it Nucl. 
Phys.} {\bf A637}, 107 (1998); {\bf A671}, 295 (2000).\vs
\bibitem{nnpap1} N. Kaiser, R. Brockmann and W. Weise, {\it Nucl. Phys. } {\bf
A625}, 758 (1997).\vs
\bibitem{nnpap2} N. Kaiser, S. Gerstend\"orfer and W. Weise, {\it Nucl. Phys.} 
{\bf A637}, 395 (1998).\vs       
\bibitem{nijmeg} M.C.M. Rentmeester, R.G.E. Timmermans, J.L. Friar and J.J. de
Swart, {\it Phys. Rev. Lett.} {\bf 82}, 4992 (1999).\vs
\bibitem{mach} R. Machleidt and D. Phillips, private communications.\vs
\bibitem{3pi} N. Kaiser, {\it Phys. Rev.} {\bf C61} 014003 (2000); {\bf C62} 
024001 (2000); {\bf C63} 044010 (2001).\vs
\bibitem{aspects} V. Bernard, N. Kaiser and Ulf-G. Mei{\ss}ner, {\it
Nucl. Phys.} {\bf A615}, 483 (1997).\vs
\bibitem{nadja} N. Fettes, Ulf-G. Mei{\ss}ner and S. Steininger, {\it Nucl. 
Phys.} {\bf A640}, 199 (1998).\vs
\bibitem{pavan} M.M. Pavan et al., {\it Physica Scripta}, {\bf T87} 65 
(2000).\vs 
\bibitem{evgeni} The same result has been independently obtained by
E. Epelbaum (unpublished).\vs 
\end{thebibliography}
\end{document}